\documentclass[prl,twocolumn,preprintnumbers,showpacs,amsmath,amssymb,superscriptaddress]{revtex4}

\usepackage{graphicx}
\usepackage{dcolumn}
\usepackage{color}
\usepackage{bm}
\usepackage{amsmath}

\usepackage{dcolumn}
\renewcommand*{\thefootnote}{\fnsymbol{footnote}}

\begin{document}

\title{Is it necessary to achieve sub-wavelength interference with correlation?}
\author{Ling-Yu Dou}
\affiliation{School of Physics, Beijing Institute of Technology, Beijing 100081, China}
\author{De-Zhong Cao}
\email{dzcao@ytu.edu.cn}
\affiliation{Department of Physics, Yantai University, Yantai 264005, China}
\author{De-Qin Xu}
\affiliation{School of Science, Tianjin University of Technology and Education, Tianjin 300222, China}
\author{Xin-Bing Song}
\email{songxinbing@bit.edu.cn}
\affiliation{School of Physics, Beijing Institute of Technology, Beijing 100081, China}
\date{\today}

\begin{abstract}
We report an  experimental demonstration of sub-wavelength interference without correlation. Typically, people can achieve sub-wavelength effect with correlation measurement no matter by using bi-photon or thermal light sources. Here we adopt a thermal light source. And we count the realizations in which the intensities of the definite symmetric points are above or below a certain threshold. The distribution of  numbers of these realizations who satisfy the restriction will show a sub-wavelength effect. With proper constrictions, positive and negative interference patterns are demonstrated.
\end{abstract}


\maketitle

Since the first demonstration of correlation imaging and interference with entangled source\cite{TBP,DVS}, correlation has been a hot topic in the area of imaging. To achieve a high qualified image, a lot of schemes are explored\cite{RSB,AG,JS,JHS,KWC,OK,FF,BMRL,XXX,DAM,SFK}. Beside imaging, interference is widely discussed with correlation method\cite{FDA,YS1,YXD,YXLW,DJSL,DGK}. Among all the correlation effects, sub-wavelength effect has attracted more attention because this effect can break the limit of diffraction\cite{MDM,GSAY,JX}, that means the resolution can be improved by a factor of 2 in theory. However, the sub-wavelength must be achieved by coincidence measurement with two-photon source or correlation measurement with thermal field, i.e. second-order correlation. Is it necessary to achieve sub-wavelength effect by second order correlation?

In this paper, we demonstrate a scheme to show sub-wavelength interference without correlation. The setup is depicted in Fig.\,1.  A He-Ne laser beam with wavelength $\lambda = 632.8 \, \rm{nm}$ impinges on a slowly rotating ground glass with a rotation frequency of $2\times 10^{-3}\,\rm{Hz}$ to form the pseudo-thermal light source. A double-slit($150\,\mu m$ width and  $300\,\mu m$ separation, center to center) is placed directly after the ground glass. The light will be recorded by a charge-coupled device (CCD, whose output is from 0 to 255) camera. In our experiment, the laser beam has a diameter of $3\,\rm{mm}$ on the ground glass. The distances from the double-slit to the CCD is $40\,\rm{cm}$. And 10000 realizations are adopted.
\begin{figure}[ht]
\centering
\includegraphics[angle=0,width=0.45\textwidth]{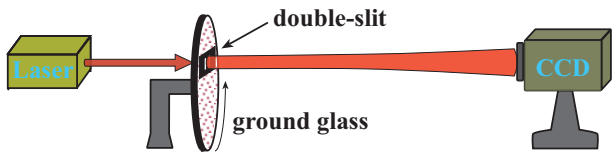}
\caption{(Color online) Experimental setup. A laser beam passes through a rotating ground glass to generate pseudo-thermal light. A double-slit is placed after the ground glass. A CCD camera will record the speckles scatted from the double-slit.}
\label{Fig. 1}
\end{figure}

If we mark the intensity of position $x$ on CCD plane as $I(x)$, and we calculate the intensity correlation between a fixed position $x_0$ and the whole plane, then a classical diffraction pattern can be observed, like
\begin{equation}\label{eq1}
g^{(2)}_{cc}(x)=\frac{\langle I(x_0)I(x)\rangle}{\langle I(x_0)\rangle \langle I(x)\rangle}
=1+\frac{\langle \Delta I(x_0) \Delta I(x)\rangle}{\langle I(x_0)\rangle \langle I(x)\rangle},
\end{equation}
where $\langle \cdot \rangle$ means ensemble average, and $\Delta I= I-\langle I \rangle$.

When the correlation is between opposite directions, i.e. $x$ and $-x$, a sub-wavelength effect appears
\begin{equation}\label{eq2}
g^{(2)}_{sub}(x)=\frac{\langle I(x)I(-x)\rangle}{\langle I(x)\rangle \langle I(-x)\rangle}
=1+\frac{\langle \Delta I(x) \Delta I(-x)\rangle}{\langle I(x)\rangle \langle I(-x)\rangle}.
\end{equation}
Because of the background in the above two correlations, the contrast has a maximum of $1/3$ in theory.

It seems widely known that sub-wavelength effect can be observed in correlation. However, we will show below that, without correlation, we still can find this effect. To avoid the complexity in statistical theory, here we choose an ideal and simple model to explain this effect qualitatively. We assume that the light passes through the double slit with equal intensities and generate random centered interference fringes as shown in Fig.\,2. The modal can be like this, for each realization there is a random phase $\phi$ in the upper slit and the diffraction fringes will shift randomly. Here we can assume that the phase $\phi$ has a distribution between 0 and $2\pi$. Apparently, we can write the fringe distribution as $I(\tilde{x},\phi)=\frac{cos(\tilde{x}+\phi)+1}{2}=cos^2(\frac{\tilde{x}+\phi}{2})$. Here $\tilde{x}=2\pi\frac{x}{\Lambda}$ is the normalized position and $\Lambda$ is the period of classical diffraction fringes. For simplicity, we normalized the maximum value of $I$ to 1. When we consider the intensities of symmetric positions ($\tilde{x}$ and $-\tilde{x}$) above some threshold value $I_{\textrm{th}}$, that is
\begin{eqnarray}\label{eq3}
&&I(\tilde{x},\phi)=cos^2(\frac{\tilde{x}+\phi}{2})\geq I_{\mathrm{th}},\nonumber\\
&&I(-\tilde{x},\phi)=cos^2(\frac{-\tilde{x}+\phi}{2})\geq I_{\mathrm{th}}.
\end{eqnarray}
This means that the probability of both the two intensities above $I_{\mathrm{th}}$ is the weight of $\phi$ which make Eq.\,(\ref{eq3}) satisfied. We know the period of the fringe of intensity is $2\pi$, and now let us consider the period of the distribution of symmetric positions. From the experimental results we can find that the periods of them are half of the period of intensity diffraction pattern. Here we give a simple explanation. When we set $\tilde{x}=\tilde{x'}+\pi$, we will have $I(\tilde{x},\phi)=cos^2(\frac{\tilde{x'}+\phi+\pi}{2})$ and $I(-\tilde{x},\phi)=cos^2(\frac{-\tilde{x'}+\phi-\pi}{2})$. If we set $\phi'=\phi+\pi$, we have $I(\tilde{x},\phi)=cos^2(\frac{\tilde{x'}+\phi'}{2})=I(\tilde{x'},\phi')$ and $I(-\tilde{x},\phi)=cos^2(\frac{-\tilde{x'}+\phi'-2\pi}{2})=I(-\tilde{x'},\phi')$. So $\pi$ is the period of the distribution for the symmetric positions. Apparently, it is half of the period for classical diffraction fringes.
\begin{figure}[ht]
\centering
\includegraphics[angle=0,width=0.45\textwidth]{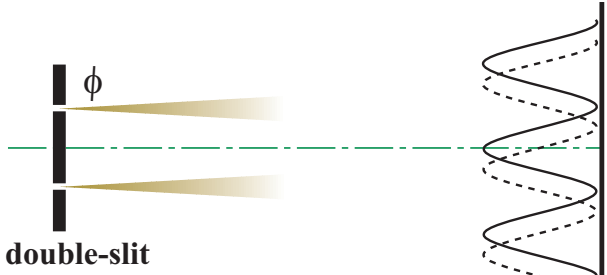}
\caption{(Color online) Sketch of the random phase interference model. The intensities from the two slits are assumed equal and the phase of light from upper slit is random.}
\label{Fig. 2}
\end{figure}
\begin{figure}[ht]
\centering
\includegraphics[angle=0,width=0.45\textwidth]{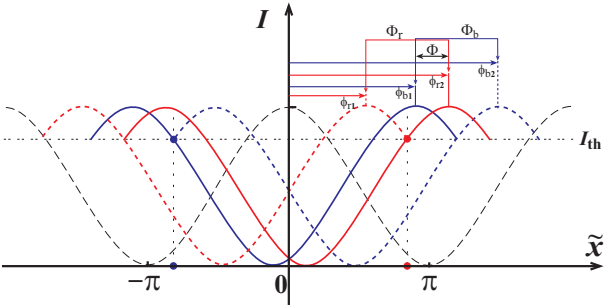}
\caption{(Color online) Interference curves for different phase differences. Black dash line is the interference curve for $\phi=0$. Red dash and real lines demonstrate the range of intensity of red spot above the threshold intensity. Blue dashed and real lines demonstrate the range of intensity of blue spot above the threshold intensity.}
\label{Fig. 3}
\end{figure}

In Fig.\,3, we give a detailed analysis when the two symmetric intensities are above the threshold intensity $I_{\mathrm{th}}$. For other situations, the analysis is similar. As shown in  Fig.\,3, the two symmetric positions $\tilde{x'}$ and $-\tilde{x'}$ are marked as red spots and blue spots, respectively. The threshold intensity $I_{\mathrm{th}}$ is marked as the horizontal dashed line. The black dashed line demonstrates the diffraction curve for $\phi=0$. When it is shift from $\phi=\phi_{r1}$ to $\phi=\phi_{r2}$, the intensity of position red spot($\tilde{x'}$) will above $I_{\mathrm{th}}$. So, $\Phi_{r}$ is the range which make the first equation in Eq.\,(\ref{eq3}) satisfied. It is not hard to find out that $\phi_{r1}=\tilde{x'}-\cos^{-1}(2I_{\mathrm{th}}-1)$ and $\phi_{r2}=\tilde{x'}+\cos^{-1}(2I_{\mathrm{th}}-1)$, where $\cos^{-1}$ is an inverse trigonometric functions. Similarly, we can find the range $\Phi_{b}$ for the blue spot position. So, the widthes of $\Phi_{r}$ and $\Phi_{b}$ are  same, i.e. $2\cos^{-1}(2I_{\mathrm{th}}-1)$, which we marked as $\Phi_{\mathrm{th}}$. The overlap between $\Phi_{r}$ and $\Phi_{b}$, marked as $\Phi$ ($0\leq\Phi\leq2\pi$), is the phase range which make Eq.\,(\ref{eq3}) satisfied. It is not hard to conclude that when $I_{\mathrm{th}}$ is increased the range of $\Phi$ will be smaller and the visibility will increase.  For this simple model, we can give an approximate expression about the visibility as 1  for $0<\Phi_{\mathrm{th}}\leq \pi$, and $\frac{2\pi-\Phi_{\mathrm{th}}}{3\Phi_{\mathrm{th}}-2\pi}$ for  $\pi<\Phi_{\mathrm{th}}\leq 2\pi$ .
However, our experimental source is definitely different from this ideal modal. This model will give a homogeneous distribution when we check the intensities of a certain pixel on the camera. But, the intensities have a similar negative exponential probability distribution, as shown in Fig.\,4.
Although the analytical formula is not precise for our experiment, the conclusion about the visibility is agree with our experimental results qualitatively as shown in Figs.\,6(a-d).

\begin{figure}[ht]
\centering
\includegraphics[angle=0,width=0.45\textwidth]{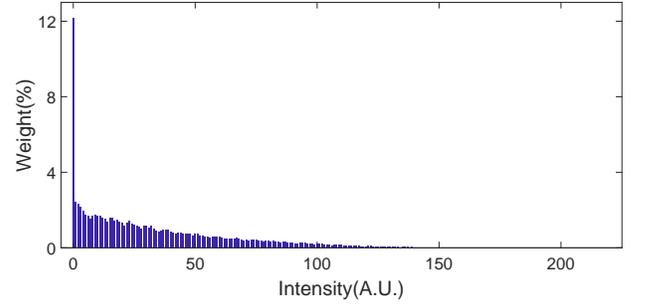}
\caption{(Color online) Intensity statistical distribution.}
\label{Fig. 4}
\end{figure}

Figure\,5 depicts out experimental results. Left column is the 2-D results and right column show the corresponding cross section results. Fig.\,5(a) is the normalized second order correlation between a fixed position (center) and the whole plane. The results, as expected, demonstrate the classical diffraction patterns. Fig.\,5(b) is the sub-wavelength correlation, and we can find that the width of the fringe is a half of the width of the fringe in Fig.\,5(a). These results are well known over ten years. The last three rows are the results when we group the realizations. We count the realizations when the intensities of symmetric positions ($x$ and $-x$) above the average intensity, and the results are shown in Fig.\,5(c). We can find that the distribution of the fringe is nearly the same as Fig.\,5(b), if we ignore the difference of amplitude. Similarly, the realizations for the intensities of $x$ and $-x$ below the average intensity are shown in Fig.\,5(d). We can notice that we also get positive fringes. This is because the intensities on peak positions are synchronism. Fig.\,5(e) shows the negative fringes when we set the intensity of $x$ bigger than the average intensity and the intensity of $-x$ smaller than the average intensity.
\begin{figure}[ht]
\centering
\includegraphics[angle=0,width=0.45\textwidth]{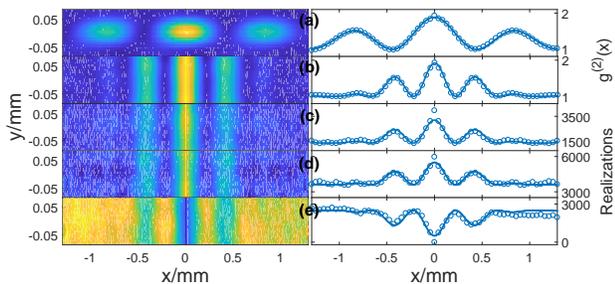}
\caption{(Color online) Experimental results. Left column is the 2-D results and right column show the corresponding cross section results. In the right column, open circles are experimental results and solid curves are theoretical fitting. (a) is the normalized second order correlation between a fixed position (center) and the whole plane. (b) is the sub-wavelength correlation. The labels of Y-axis for (a) and (b) are marked as $g^{(2)}(x)$. (c) is the realizations when intensities of symmetric positions ($x$ and $-x$) above the average intensity. Similarly, the realizations for the intensities of $x$ and $-x$ below the average intensity are shown in (d). (e) shows a negative fringe when we set the intensity of $x$ bigger than the average intensity and the intensity of $-x$ smaller than the average intensity. The labels of Y-axis for (c)-(e) are marked as Realizations.}
\label{Fig. 5}
\end{figure}

We have demonstrated the sub-wavelength pattern without correlation. Next we will explore the visibility of the fringes qualitatively. In the experiment, we take 10000 realizations and the average intensity is about 34 (arbitrary unit). The details are shown in figure 6. Figs.\,6(a)-(d) are the results for the situation of intensities of symmetric positions greater than 20, 34, 60 and 100, respectively. And the visibility are 0.34, 0.49, 0.72 and 0.94, respectively. Here the visibility is defined as $(r_{max}-r_{min})/(r_{max}+r_{min})$, and $r$ is the realizations. Similarly, when the intensities of symmetric positions are smaller than 20, 34, 60 and 100, the corresponding curves are shown in Figs.\,6(e)-(h), and the visibility are 0.39, 0.25, 0.12 and 0.033, respectively. We can find that when the threshold is higher the visibility becomes greater for intensities of symmetric positions greater than the thresholds. Inversely for the situation of intensities of symmetric positions smaller than the thresholds, when the threshold is higher the visibility is lower.

\begin{figure}[ht]
\centering
\includegraphics[angle=0,width=0.45\textwidth]{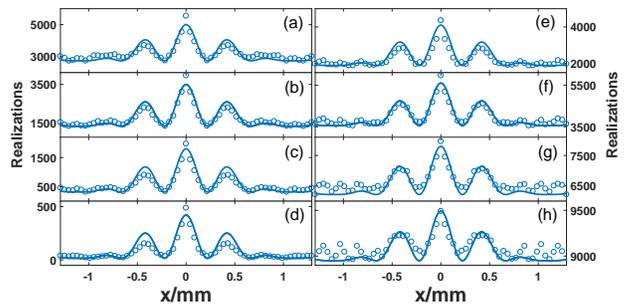}
\caption{(Color online) Experimental results . Left column and right column show the results for the situation of intensities of symmetric positions bigger and smaller than the thresholds, respectively. From top to bottom, the thresholds are set to 20, 34, 60 and 100, respectively.
The visibility are 0.34, 0.49, 0.72 and 0.94 for (a)-(d), respectively. The visibility are 0.39, 0.25, 0.12 and 0.033 for (e)-(h), respectively.}
\label{Fig. 6}
\end{figure}

In summary, we experimentally study the sub-wavelength effect without correlation. We sort the realizations according the intensities of the symmetrical positions above or bellow some threshold, and we can get positive and also negative patterns. We also notice that Wu's and Shih's groups finished positive and negative imaging by sort the speckles patterns and accumulating them \cite{LAW,REM,KHL}. However, their results are still intensity distribution. Recently, Wu's group demonstrated an experiment about sub-wavelength diffraction with thermal light\cite{MJS}. Their result is based on the second-order correlation. Until now, to our knowledge, all the sub-wavelength effect are based on correlation, but our results show that correlation calculation is not necessary for sub-wavelength effect. We wish our results can deepen the understanding of sub-wavelength effect, and find some potential applications in image processing and signal processing.

\begin{acknowledgments}
This work was supported by the National Natural Science Foundation of China (Grant No. 11304016).
\end{acknowledgments}

\bibliographystyle{apsrev}

\end{document}